# Magnetic dynamics of single domain Ni nanoparticles

G. F. Goya, F.C. Fonseca, and R. F. Jardim
*Instituto de Física, Universidade de São Paulo, CP 66318, 05315-970, São Paulo, Brazil*

R. Muccillo
*Centro Multidisciplinar de Desenvolvimento de Materiais Cerâmicos CMDMC, CCTM-Instituto de Pesquisas Energéticas e Nucleares CP 11049, 05422-970, São Paulo, SP, Brazil*

N. L. V. Carreño, E. Longo, and E.R. Leite
*Centro Multidisciplinar de Desenvolvimento de Materiais Cerâmicos CMDMC, Departamento de Química, Universidade Federal de São Carlos CP 676, 13560-905, São Carlos, SP, Brazil*

The dynamic magnetic properties of Ni nanoparticles diluted in an amorphous $SiO_2$ matrix prepared from a modified sol-gel method have been studied by the frequency $f$ dependence of the ac magnetic susceptibility $\chi(T)$. For samples with similar average radii ~ 3-4 nm, an increase of the blocking temperature from $T_B$ ~ 20 to ~ 40 K was observed for Ni concentrations of ~ 1.5 and 5 wt.%, respectively, assigned to the effects of dipolar interactions. Both the in-phase $\chi'(T)$ and the out-of-phase $\chi''(T)$ maxima follow the predictions of the thermally activated Néel-Arrhenius model. The effective magnetic anisotropy constant $K_{eff}$ inferred from $\chi''(T)$ versus $f$ data for the 1.5 wt.% Ni sample is close to the value of the magnetocrystalline anisotropy of bulk Ni, suggesting that surface effects are negligible in the present samples. In addition, the contribution from dipolar interactions to the total anisotropy energy $E_a$ in specimens with 5 wt.% Ni was found to be comparable to the intrinsic magnetocrystalline anisotropy barrier.

The dynamics of ferromagnetic nanoparticles with different interaction strengths has been widely studied in recent years.[1,2] The model describing the magnetic behavior of a system of monodispersed and noninteracting single-domain particles proposed by Néel[3] has been successfully tested by numerous experiments with increasing sophistication, as the delicate series of works that have recently confirmed its applicability at the single-particle level.[4]

For a single-domain particle, the energy barrier between magnetic states may be considered to be proportional to the particle volume V. In the case of uniaxial anisotropy, the anisotropy energy $E_a$ in the absence of external magnetic field is described by $E_a = K_{eff} V \sin^2\theta$, where $K_{eff}$ is an effective magnetic anisotropy constant and $\theta$ is the angle between the magnetic moment of the particle and its easy magnetization axis. On the other hand, the dynamic response of such particles to an alternating external magnetic field is determined by the measuring time $\tau_m$ of each experimental technique. Since reversion of the magnetic moments over the anisotropy energy barrier $E_a$ is assisted by thermal phonons, the relaxation time $\tau$ of each magnetic particle exhibits an exponential dependence on temperature characterized by an Néel-Arrhenius law

$$\tau = \tau_0 \exp\left[\frac{E_a}{k_B T}\right] \qquad (1)$$

where $f_0 = \tau_0^{-1}$ is an attempt frequency. Typical values for $\tau_0$ are in the $10^{-9}$ - $10^{-11}$ s range for superparamagnetic (SPM) systems.

When an *ensemble* of single-domain magnetic particles is considered, the above description is still valid provided that the particles are non interacting.

However, as the concentration of the magnetic phase increases, interparticle interactions alter the single-particle energy barrier, and concurrent effects involving dipolar interactions, particle size distribution, and aggregation make the application of Eq. (1) not obvious. To better understand how dipolar interactions affect the SPM relaxation rates it is therefore desirable to prepare samples near the infinite-dilution limit of the magnetic phase, settling the single-particle properties of a specific magnetic system, and then gradually increase the particle density. In this work we have used the above approach to study the dynamics of magnetic properties in high-quality Ni nanoparticles. The samples were prepared by a modified sol-gel technique and characterized by ac magnetic susceptibility $\chi(T)$ measurements as a function of temperature, applied field, and excitation frequency.

Nanocomposites of Ni:SiO$_2$ were synthesized by using tetraethylorthosilicate (TEOS), citric acid, and nickel (II) nitrate. The citric acid was dissolved in ethanol and the TEOS and the nickel nitrate were added together and mixed for homogenization at room temperature. After the polymerizing reaction adding ethylene glycol, the solid resin was heated at 300 °C, ground in a ball mill, and then pyrolyzed at 500 °C. Further details of the method employed can be found elsewhere.[5] In the present work we will concentrate our discussion in two samples having ~1.5- and 5-wt.% Ni which will be referred as *S*1 and *S*2, respectively. The structure and morphology of the magnetic powders were examined by transmission electron microscopy with a 200-kV, high-resolution transmission microscope. Magnetization and ac magnetic susceptibility measurements were performed in a commercial SQUID magnetometer both in zero-field-cooling (ZFC) and field-cooling (FC) modes, between 1.8 K < T < 300 K and under applied fields up to 7 T. The frequency dependence of both in-phase $\chi'(T)$ and out-of-phase $\chi''(T)$ components of the ac magnetic susceptibility was measured by using an excitation field of 2 Oe and driving frequencies between 20 mHz $< f <$ 1.5 kHz.

We have previously characterized these two samples of Ni nanoparticles embedded in $SiO_2$ by several techniques and have observed some features which are summarized as follows: (1) a log-normal distribution of particle sizes with average radius close to ~ 3-4 nm for both samples; (2) the occurrence of a SPM behavior above $T_B >$ 20 K and 40 K, for samples S1 and S2, respectively; (3) a nearly spherical morphology for both samples; and (4) the absence of a shell-core NiO-Ni morphology, where an antiferromagnetic layer of NiO (shell) surrounds the ferromagnetic Ni (core) particles.[6]

Turning now to the dynamics of the magnetic particle systems, Fig. 1 displays the temperature dependence of $\chi'(T)$ and $\chi''(T)$ of the more diluted sample S1 and for different frequencies $f$. The data for both components $\chi'(T)$ and $\chi''(T)$ exhibit the expected behavior of a SPM system, i. e., the occurrence of a maximum in temperature for both $\chi'(T)$ and $\chi''(T)$ components, and a shift of this maximum towards higher temperatures with increasing frequency. The freezing of the magnetic moments from the SPM to a blocked state occurs at the blocking temperature, $T_B$, at which the relaxation time $\tau$ of the Ni nanoparticles is equal to the experimental time window $\tau_e = 1/f$ of the ac measurement, $T_B = \beta E_a / k_B \ln(1/f \tau_0)$.[7] In this expression $\beta$ represents the effect of the particle size distribution g(D), being $\beta = 1$ for a monodispersed sample (i.e., a delta $g(D) = \delta(D-D_0)$ size distribution). However, spin-glass systems also display features similar to the ones described above and it seems convenient to classify first our Ni nanoparticles. An empirical and model-independent criterion used for classifying a transition to a frozen state is the relative shift of the temperature of the maximum in χ"(T), $T_m$, with the measuring frequency $f$ as

$$\Phi = \frac{\Delta T_m}{T_m \Delta \log_{10}(f)} \qquad (2)$$

where $\Delta T_m$ is the difference between $T_m$ measured in the $\Delta\log_{10}(f)$ frequency interval.

Experimentally, the $\Phi$ values found for SPM systems are in the range ~0.10-0.13, whereas a much smaller dependence of $T_m$ with $f$ is observed in spin glasses ($\Phi \sim 5\times10^{-3} - 5\times10^{-2}$).[2,8] Therefore, Eq. (2) provides a model-independent classification of the kind of freezing transition. However, it is well known that intermediate situations (0.001< $\Phi$ < 0.05) are often reported, usually related to non-diluted particulate systems.[2,9] Our calculated values of $\Phi = 0.12$ and 0.13 for samples S1 and S2, respectively, show unambiguously that the shift in $T_m$ with increasing $f$ corresponds to a thermally activated Néel-Arrhenius model for superparamagnets.

This behavior was confirmed by the linear dependence of $\ln[\tau]$ versus $1/T_B$ shown in Fig. 2 for both samples. It can be further seen that both curves are fitted very well by using Eq. 1 and show the same extrapolated value of $\tau_0 = 8\times10^{-10}$ s, consistent with a SPM system. The frequency dependence of $T_B$ in Eq. 1 is determined by the effective activation energy barrier $E_a$. Contributions to $E_a$ can originate from intrinsic anisotropies of the particles (shape, magnetocrystalline, or stress ansotropies) or interparticle interactions (dipolar or exchange). Inasmuch as these two mechanisms contribute to modify the energy barrier, it is usually quite difficult to separate both kind of effects.

The values of $K_{eff}$ of our samples were extracted from the activation energies by using the average particle radii from TEM data ($r_m = 4.2$ and 3.3 nm for S1 and S2, respectively) and then compared to the first-order magnetocrystalline anisotropy constant at low temperature $K_1^{bulk} = -8 \times 10^5$ erg/cm$^3$ of bulk Ni.[10] For the present case, with cubic anisotropy and $K_1 < 0$, the effective (uniaxial) anisotropy is related to $K_1$ through the relation $K_{eff} = K_1/12$.[7] Therefore, from the $K_{eff} = 1.3 \times 10^5$ erg/cm$^3$ value obtained for S1 a magnetocrystalline anisotropy of $K_1 = 15 \times 10^5$ erg/cm$^3$ is extracted, which is only twice the value of $K_1^{bulk}$. If shape anisotropy is assumed as the only source of anisotropy, a small deviation from spherical shape (e.g., to prolate spheroidal) to an axis ratio c/a ~ 1.2 would be enough to explain the calculated value of $K_{eff}$. On the other hand, it is useful to relate $K_{eff} = 1.3 \times 10^5$ erg/cm$^3$ with the expected coercive field for purely magnetocrystalline anisotropy of spherical particles $H_C = 2K_{eff}/M_S \approx 500$ Oe, a value in excellent agreement with $H_C$ ~ 520 Oe obtained from hysteresis curves at low temperatures.[6] Therefore, the above data suggest that these Ni particles have indeed nearly spherical shape, with intrinsic magnetic anisotropy close to the Ni (fcc) bulk value.

Returning to the curves shown in Fig. 2, it is also clear that the energy barriers increase with increasing Ni content, as inferred from the larger slope of ln$\tau$ vs. $T_B^{-1}$ curves. Such an increase in $E_a$ can not be related to a larger average volume of the Ni particles in sample S2 since both radius distributions have similar mean values. Actually, the average radius $r_m$ extracted from the log-normal distribution of sample S2 is slightly *smaller* than for sample S1.[6] Similarly, from our previous discussion regarding the value of $K_{eff}$ obtained for sample S1, a significant contribution from surface effects to $E_a$ in sample S2 seems to be unlikely. Therefore, the increase of the effective energy barrier for the more concentrated sample should be related to dipolar interactions.

Following this discussion, we have estimated this dipolar contribution to the total energy by comparing the values of $E_a$ for both samples S1 and S2. Based on the similar volume distributions from TEM images, we assume that Ni nanoparticles in both samples have similar intrinsic anisotropies. Within this context, the only effect of increasing concentration is thus to add a dipolar term $E_{dip}$ to the effective activation energy $E_a$. Following Luis et al.[11] we have used a modified Arrhenius-Néel expression for the relaxation time including the contribution of the dipolar energy as $\tau = \tau_0 \exp\{(U_0 + E_{dip})/k_B T\}$, where $U_0$ is the single-particle energy barrier. From this relationship, we can write

$$\ln\left(\frac{\tau_2}{\tau_1}\right) = \frac{E_{dip}}{k_B T} \tag{3}$$

where $\tau_1$ and $\tau_2$ are the relaxation times of samples S1 and S2, respectively. From the $E_a$ values fitted for samples S1 and S2, we obtained $E_{dip} = 247$ K. This value is comparable to $U_0 = 282$ K for single-domain and isolated Ni nanoparticles as estimated from S1 sample.

In conclusion, we have studied the dynamics of ferromagnetic Ni nanoparticles with similar radius distributions and different concentrations *via* ac magnetic susceptibility measurements. The general behavior of these nanoparticles is well described by the Néel-Arrhenius model for single-domain, noninteracting particles. For the more diluted sample with 1.5 wt.% Ni, the estimated magnetic anisotropy of the particles was similar to the value of the magnetocrystalline anisotropy for bulk (fcc) Ni, suggesting that both shape and surface anisotropies are negligible. For the more concentrated sample with 5 wt.% Ni, the increase of the energy barrier $E_a$ could be described by an additional contribution $E_{dip}$ coming from dipolar interactions. We estimated $E_{dip} \approx 247$ K, a value comparable to the intrinsic magnetic anisotropy $U_0 \approx 282$ K for single-domain nanoparticles.

This work was supported in part by the Brazilian agency Fundação de Amparo à Pesquisa do Estado de São Paulo (FAPESP) under Grant Nos. 99/10798-0; 01/02598-3; 01/04231-0; 98/14324-0; and the Conselho Nacional de Desenvolvimento Científico e Tecnológico (CNPq) under Grant Nos. 300569/00-9 and 304647/90-0.

Figure Captions

Figure 1. Temperature dependence of the real component $\chi'(T)$ of the magnetic susceptibility for 1.5 wt.% Ni (sample S1) at different excitation frequencies. Inset: Imaginary part $\chi''(T)$ for the same sample shown in an expanded T-scale. The data were taken with an external magnetic field H of 50 Oe.

Figure 2. Arrhenius plots of the relaxation time $\tau$ vs. blocking temperature $T_B$ obtained from the imaginary component $\chi''(T)$ of the ac magnetic susceptibility. Dashed lines are the best fit using Eq. (1) with a single $\tau_0$ value and $E_a(S1) = 282$ K, $E_a(S2) = 529$ K.

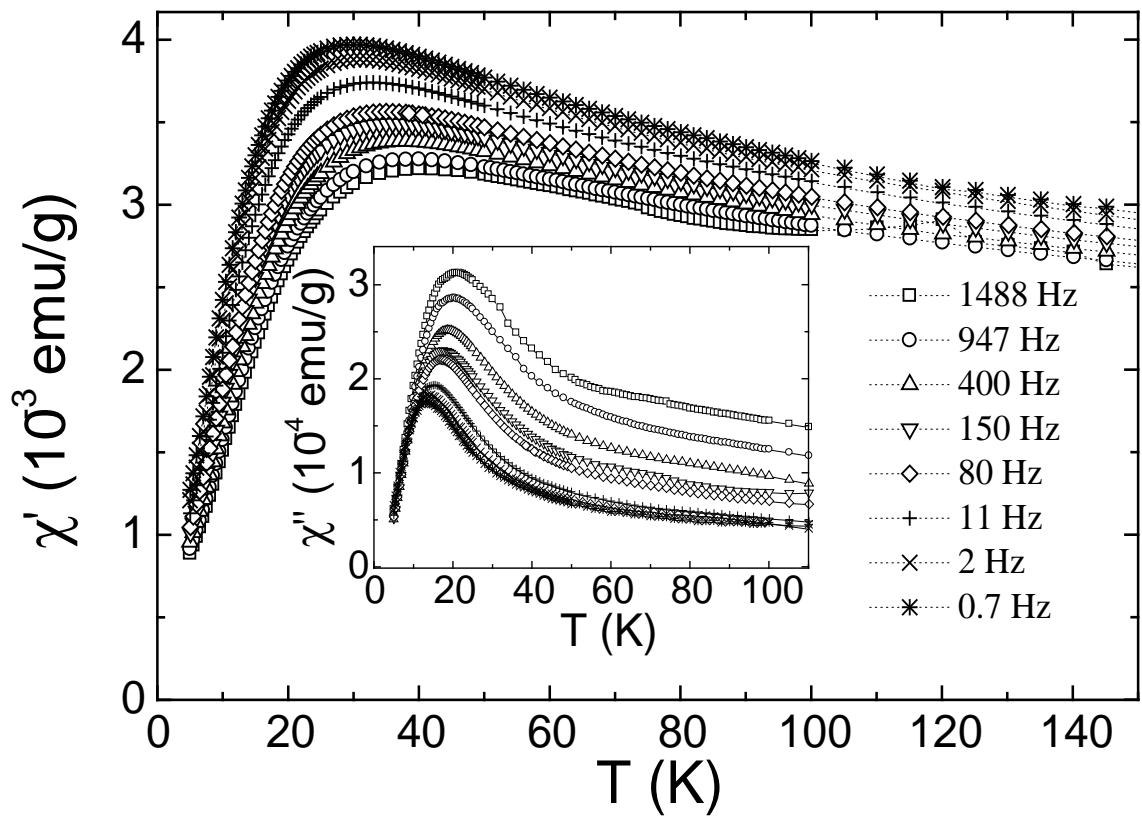

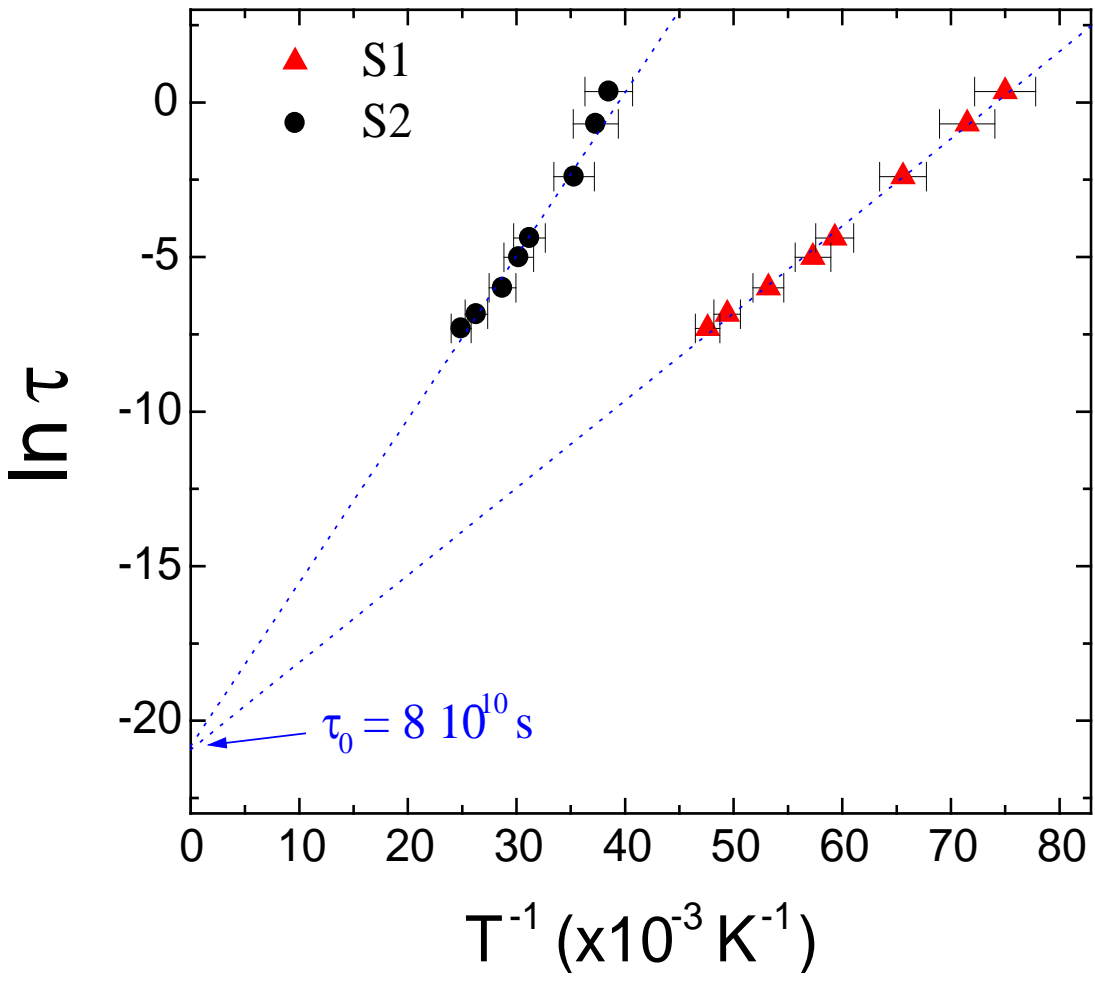

Figure 2
Gova et al